# Transfer learning for materials informatics using crystal graph convolutional neural network

Joohwi Lee[a,*] and Ryoji Asahi[a,+]

[a] *Toyota Central R&D Laboratories, Inc., Nagakute, Aichi 480-1192, Japan*

[*] Correspondence e-mail : j-lee@mosk.tytlabs.co.jp (J. Lee)

[+] Present address: *Institute of Materials Innovation, Nagoya University, Furo-cho, Chikusa-ku, Nagoya, Aichi 464-8603, Japan*

**Abstract**

For successful applications of machine learning in materials informatics, it is necessary to overcome the inaccuracy of predictions ascribed to insufficient amount of data. In this study, we propose a transfer learning using a crystal graph convolutional neural network (TL-CGCNN). Herein, TL-CGCNN is pretrained with big data such as formation energies for crystal structures, and then used for predicting target properties with relatively small data. We confirm that TL-CGCNN can improve predictions of various properties such as bulk moduli, dielectric constants, and quasiparticle band gaps, which are computationally demanding, to construct big data for materials. Moreover, we quantitatively observe that the prediction of properties in target models via TL-CGCNN becomes more accurate with an increase in size of training dataset in pretrained models. Finally, we confirm that TL-CGCNN is superior to other regression methods in the predictions of target properties, which suffer from small amount of data. Therefore, we conclude that TL-CGCNN is promising along with compiling big data for materials that are easy to accumulate and relevant to the target properties.

# 1. Introduction

To save time and human efforts, machine learning (ML) [1-6] has been widely applied to the materials informatics (MI) field, particularly for new material exploration. An important challenge in ML-based material exploration and design is the development of a generic descriptor (representation) that can be flexibly applied to predictions of multiple target properties. Because material properties depend on the crystal structure and constituent chemical elements, descriptors based on the structural features, such as Coulomb matrix [7, 8], smooth overlap of atomic positions (SOAP) [9, 10], and reciprocal 3D voxel space (R3DVS) [11], have been developed.

Recently, Xie and Grossman introduced a crystal graph convolutional neural network (CGCNN) [12, 13]. As an input to a classification or regression model based on the CGCNN, only a crystal structure is necessary. The CGCNN constructs crystal graphs from crystal structures and predicts the target property using a deep neural network architecture. Numerous researchers have modified and improved on graph-based neural networks owing to their simplicity and flexibility. Park and Wolverton incorporated information of the Voronoi tessellated crystal structure into the CGCNN to encode three body correlations [14]. Sanyal *et al.* augmented the CGCNN with a multitask learning to predict multiple material properties simultaneously [15]. Chen *et al.* developed a MatErials Graph Network (MEGNet) that considers not only crystal structures but also molecules with additional attributes [16]. Karamad *et al.* implemented atomic orbital interaction features [17] and encoder-decoder networks in the CGCNN [18].

Another important challenge in ML-based material exploration and design is to overcome the unsatisfactory accuracy of predictive models because of insufficient amount of data [19]. Compared to other research fields such as computer vision [20] and natural language processing [21] which use big data comprising millions of data items [22], functional or practical properties in MI have not been substantially accumulated as their evaluation is computationally or experimentally demanding. Transfer learning (TL) [23] is a useful method for improving predictions suffering from insufficient amount of data.

TL is used to provide knowledge to other predictive models with different tasks (target variables). Herein, we focus on applications of TL for a target model with less amount of data by loading a pretrained model constructed with an enormous amount of data. In MI, studies have been conducted using TL. To classify Li superionic conductors, Cubuk *et al.* [24] constructed a large amount of labeled data based on crystal structure-based descriptors, and then reconstructed predictive models based on chemical element-based descriptors using a support vector machine. Yamada *et al*. [25] developed a pretrained model library called XenonPy.MDL [26] with big data for a multilayer neural network based on fingerprint descriptors [27], and it showed an improved prediction of organic or inorganic material properties by TL using the neural network. Jha *et al*. [28] demonstrated improved predictions of formation energies ($\Delta E_f$) for smaller theoretical or experimental databases by TL using the ElemNet multilayer deep neural network [29], which uses elemental composition descriptors, and pretrained models for $\Delta E_f$ with a larger theoretical database [30, 31].

Indeed, as deep convolutional neural networks become the mainstream in the computer vision field [23, 32-35], TL is being increasingly used. The convolutional layers work as feature maps that can detect basic features such as vertical and horizontal lines, which are basic elements for recognizing objects. Li *et al*. [36] applied TL for microstructure reconstruction using deep convolutional neural networks pretrained with image big data such as ImageNet [22]. Based on the inspiration, CGCNN, on account of its neural network architecture, may have the potential for effective TL in MI.

In this study, we investigate TL using CGCNN (TL-CGCNN). Additionally, we quantitatively confirm that the predictions of basic material properties such as the Kohn‑Sham band gap ($E_g$) and $\Delta E_f$ can be improved via TL-CGCNN. To confirm the flexible application of TL, we investigate whether TL-CGCNN can improve predictions of properties such as the bulk modulus ($K_{VRH}$) [37], dielectric constant ($\varepsilon_r$) [38], and quasiparticle band gap (GW-$E_g$) [39], which usually suffer from small amount of data and heavy computational costs. In addition, we discuss issues relevant to TL-CGCNN and its prospects.

**2. Method**

*2.1. CGCNN*

First, we review the background knowledge regarding the CGCNN constructed by Xie and Grossman [12, 13]. Figure 1 presents an overview of the CGCNN. The CGCNN involves the construction of graphs based on crystal structures and a deep neural network architecture including embedding, convolutional, pooling, and fully-connected (FC) layers.

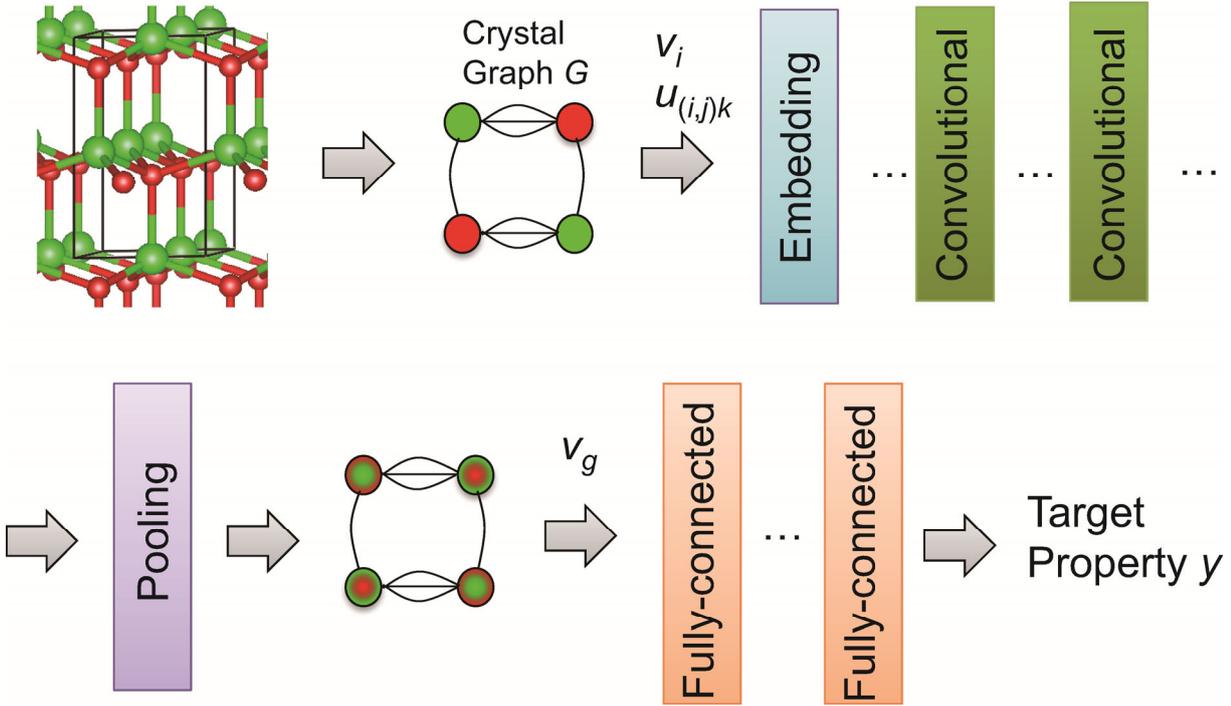

**Fig. 1.** Overview of the CGCNN. It predicts a target property with an input of the crystal structure, which is transformed into a crystal graph. Green and red atoms in the crystal structure correspond to the green and red nodes that are displayed as circles in the graph, respectively. The chemical bonds in the crystal structure correspond to the edges in the graph. The CGCNN architecture consists of embedding, convolutional, pooling, and FC layers.

The crystal structure is transformed into a multigraph $G$. Nodes and edges in the multigraph $G$ indicate atoms and chemical bonds between the atoms, respectively. The multigraph $G$ consists of a set of atoms in the crystal structure, undirected edges, atom features, and bond features. A bond feature vector ($u_{(i,j)k}$) denotes the $k$-th bond between atoms $i$ and $j$. An atom feature vector ($v_i$) contains the properties of the $i$-th atom. The information of atoms and bonds in the multigraph $G$ is prepared as discrete representations that consist of binary digits for describing constituent elements and bonds, such as group and periodic number of atoms, and atom distance. The discrete representations are transformed into continuous representations in the embedding layer. Thereafter, the continuous representations are input to the convolutional layer.

In the ($t$+1)-th convolutional layer,

$$v_i^{(t+1)} = g\left[\left(\sum_{j,k} v_j^{(t)} \oplus u_{(i,j)k}\right) W_c^{(t)} + v_i^{(t)} W_s^{(t)} + b^{(t)}\right], \tag{1}$$

where $\oplus$ denotes the concatenation of the atom and bond feature vectors of the neighboring atoms of the $i$-th atom. $W_c^{(t)}$, $W_s^{(t)}$, and $b^{(t)}$ represent the optimized parameters in the neural network, namely, the convolution weight matrix, self-weight matrix, and bias of the *t*-th layer, respectively. $g(\cdot)$ denotes a non-linear activation function between layers, such as a softplus function [40].

However, because all neighbors in the convolution formulation share the weight matrix, different interactions of neighbors are not distinguished. Therefore, a standard edge-gating technique [41] can be applied for the formulation of a new type of convolution as follows: A neighbor feature vector $z_{(i,j)k}$ is used as,

$$\boldsymbol{z}^{(t)}_{(i,j)k} = \boldsymbol{v}^{(t)}_i \oplus \boldsymbol{v}^{(t)}_j \oplus \boldsymbol{u}_{(i,j)k}, \quad (2)$$

and then,

$$\boldsymbol{v}^{(t+1)}_i = \boldsymbol{v}^{(t)}_i + \sum_{j,k} \sigma(\boldsymbol{z}^{(t)}_{(i,j)k}\boldsymbol{W}^{(t)}_c + \boldsymbol{b}^{(t)}_c) \circ g(\boldsymbol{z}^{(t)}_{(i,j)k}\boldsymbol{W}^{(t)}_s + \boldsymbol{b}^{(t)}_s), \quad (3)$$

where $\circ$ denotes element-wise multiplication (Hadamard product), and $\sigma$ denotes a sigmoid function. Xie and Grossman [12] reported that the latter convolution formulation (equation 3) showed better predictions compared to the former (equation 1).

The atom feature vectors output by the convolutional layers are inserted into the pooling layer. Herein, an average pooling [42] is used. Thereafter, a crystal feature vector ($v_g$) can be obtained by,

$$\boldsymbol{v}_g = \frac{1}{N}\sum_i \boldsymbol{v}_i, \quad (4)$$

where $N$ is the number of atoms in the crystal graph.

Finally, $v_g$ is used as input to a network of FC layers with non-linearities that learn to predict a target property value, $y$, for the crystal. The predicted values can be obtained as follows:

$$\hat{y} = f(\boldsymbol{v}_g \boldsymbol{W}_g + \boldsymbol{b}_g), \quad (5)$$

where $\boldsymbol{W}_g$, $\boldsymbol{b}_g$, and $f(\cdot)$ are the weight matrix, bias, and non-linear function of the FC layers, respectively.

The $W$ and $b$ parameters are optimized to minimize the loss function, which is defined as the mean absolute difference between the predicted and target values for regressions, by a backpropagation procedure. Herein, stochastic gradient descent is employed as an optimizer [43].

*2.2. TL-CGCNN*

Figure 2 presents a comparison of the concept of a conventional ML and a TL based on a convolutional neural network. For a conventional ML, we prepare each predictive model separately when we have different target variables. However, when we construct a "pretrained" (source) predictive model using big data based on a convolutional neural network in advance, an enormous number of parameters (weight and bias) in the convolutional layer can save the features of the big data. The parameters in the convolutional layer can be transferred to the target predictive model, which usually has a smaller amount of training data, as "knowledge".

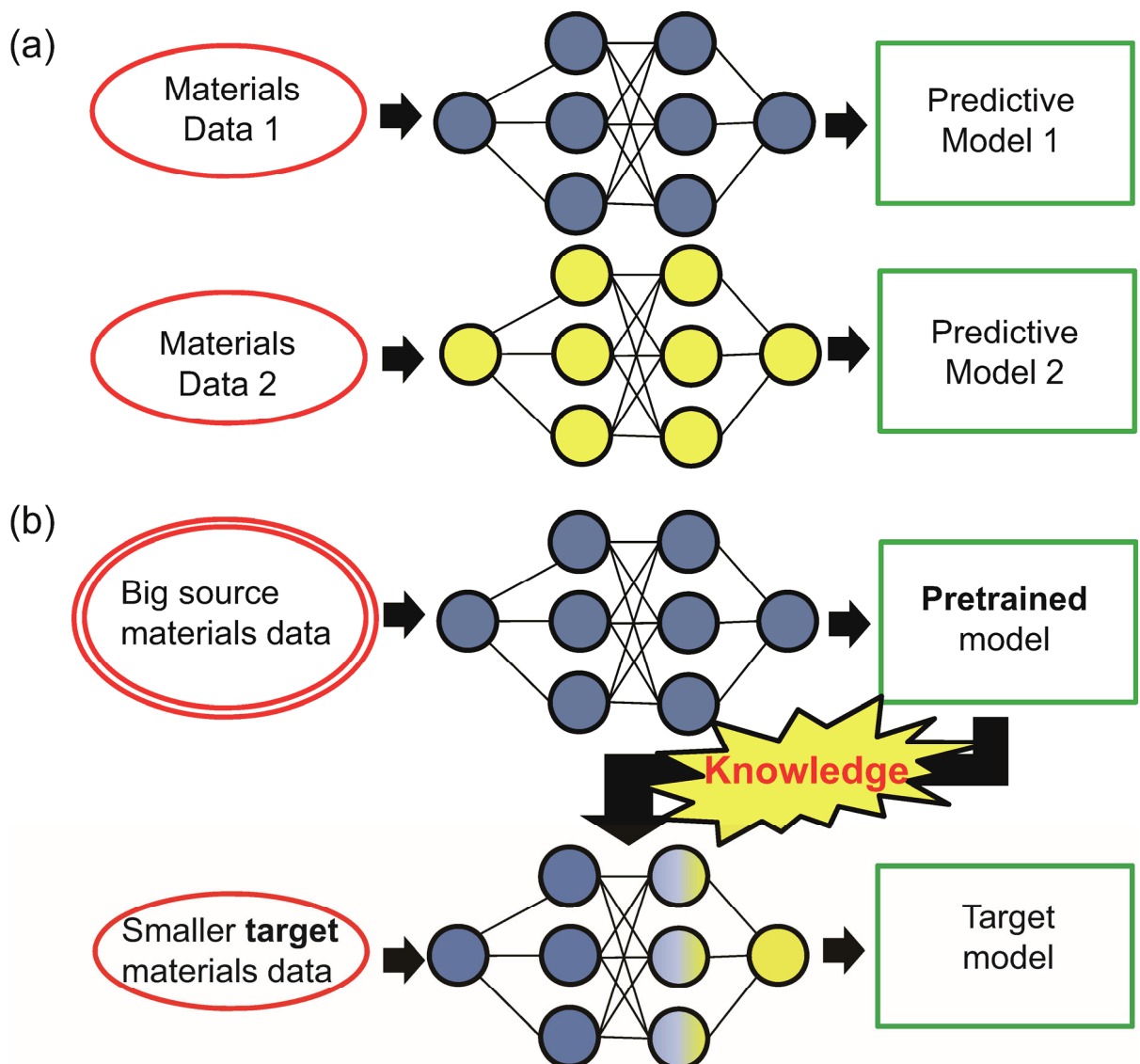

**Fig. 2.** Overview of TL. Comparison of (a) a conventional ML and (b) TL. For conventional ML, we prepare different predictive models separately when we have different tasks. However, when using TL, the knowledge (parameters, namely, weights and biases) accumulated in the pretrained model with big data can be transferred to the target predictive model.

There are two common methods in TL for treating the loaded parameters for convolutional neural networks. The first method is "fine tuning." Herein, the parameters of the entire layers of the pretrained model are initially loaded. Thereafter, all the parameters are optimized via backpropagation. This method is expected to have better optimization compared to when optimization is started with randomly generated initial parameters. To avoid losing the information of the pretrained model, it is recommended to select an adequate learning rate. The second method is "layer freezing." This method fixes the parameters of the earlier layers without additional optimizations by backpropagation. To save the information from the pretrained model, the optimization of parameters is only performed for the latter layers. In this study, we mainly used fine-tuning for TL-CGCNN.

In addition, to match the scales of target properties between the pretrained and target models in this study, tanh-normalization [44] was used for the predictive models as follows:

$$y' = \frac{1}{2}\left\{\tanh\left(\frac{0.01(y-\mu)}{s}\right) + 1\right\}, \tag{6}$$

where tanh, $\mu$, and $s$ are the hypertangent function, and mean and standard deviation of the target variable, $y$, respectively. After the predictive model is optimized, it is denormalized when the prediction error is summarized.

*2.3. Data and evaluation of predictive model*

Crystal structures and their corresponding $E_g$ and $\Delta E_f$ (defined with respect to the most stable state of each elemental standard state) have been compiled from the first-principles calculations database, Materials Project Database (MPD) [45]. They were calculated based on the exchange-correlation function of generalized gradient approximation (GGA) parameterized by Perdew-Burke-Ernzerhof (PBE) form [46] in the Vienna *Ab-initio* Simulation Package (VASP) [47, 48]. The on-site Coulomb interactions were treated using GGA+U in case transition metals were included in the materials [49]. The effective U values for the transition metals are summarized in Supplementary Table S3. We used 118286 computed data samples (assessed on November 11th, 2019) that are available in MPD to construct crystal graphs with sufficient connectivity between atoms. Within a cutoff radius of 8 Å [13], twelve nearest neighboring atoms were considered to construct the crystal graphs [12]. Crystal graphs with insufficient connectivity, which do not have twelve neighboring atoms within the cutoff radius, were not included in the data. He, Ar, and Ne were also not included in the data. Among them, 59923 data samples were used as nonmetallic (NM) materials under the condition, computed $E_g > 0.1$ eV. The CGCNN was used to predict the $\Delta E_f$ value with the input of only crystal structures for all materials. For these NM materials, the CGCNN was used to predict the $E_g$ as well as the $\Delta E_f$ value.

In this study, large sizes of datasets, namely, 10000, 54000, and 113000 are written as 10k, 54k, and 113k, respectively. The ratio of size of training, validation, and test datasets was set to 4:1:1 for data selection of the CGCNN. However, the sizes of the validation and test datasets were fixed as 2500 when the size of the training dataset was greater than 10k. Weights and biases were optimized to minimize the loss function by using training data. Validation data were used to find the best epoch to stop the backpropagation with the minimum loss function, which indicates the threshold point of overfitting. Test data that did not participate in the modeling were used to evaluate the prediction error. In this study, the mean absolute error (MAE) of the test data was used to evaluate the prediction error. The prediction errors were evaluated by averaging over several trials with different datasets. When the size of the training dataset was 500, greater than 500 and less than or equal to 10k, and greater than 10k, the number of trials was set to 10, 5, and 2, respectively. We used random selections of training/validation/test data based on random seeds. Therefore, although training/validation/test data vary among different trials, the same type of data could be used for comparisons of various predictive models when the same amount of data is used. The maximum iteration of epochs was mostly set to 500, ensuring that the minimum MAE of the validation data was already obtained at an epoch less than 500. When this was not satisfied, we increased the maximum iteration of epochs until we obtained the minimum MAE of the validation data.

TL-CGCNNs were employed to predict $E_g$ or $\Delta E_f$ with less data using the pretrained CGCNN models that were constructed to predict $E_g$ or $\Delta E_f$ with 10k, 54k, and 113k (only available for $\Delta E_f$) training data. For the target models, the ratio of the size of training, validation, and test datasets was also set to 4:1:1.

To investigate the effect of the TL-CGCNN on more practical cases that suffer from insufficient amount of data, we used the target properties of $K_{VRH}$, $\varepsilon_r$, and GW-$E_g$. The $K_{VRH}$ [37] and $\varepsilon_r$ [38] values were obtained from MPD. For the $K_{VRH}$, we prepared 12990 data samples with values ranging from 2 GPa (Cs metal) to 436 GPa (diamonds), and also obtained by the Voigt-Reuss-Hill approximation [50]. For the $\varepsilon_r$, we prepared 4182 NM data samples (with a condition of $E_g > 0.1$ eV). The $\varepsilon_r$ value is prepared with an assumption to be a single scalar polycrystalline value converted from a dielectric tensor [38]. The GW-$E_g$ values were obtained from a previous study [39]. To obtain the GW-$E_g$ values, the $G_0W_0$ calculations [51] with Heyd-Scuseria-Ernzerhof hybrid functional ($G_0W_0$@HSE06) [52] were used for 270 binary and ternary materials (inorganic compounds) with optimized crystal structures by PBE-sol functional [53].

Supplementary Table S1 summarizes the main hyperparameters that were used for the CGCNN models in this study. The default values of other hyperparameters can be referred to in the original code [13]. Among various hyperparameters, the length of the feature vector of the FC layer and the initial learning rate of the target model were reduced when we used TL-CGCNNs.

We compared the prediction of the TL-CGCNN with other widely used regression methods for various predictive models. Partial least squares (PLS) [54], least absolute shrinkage and selection operator (LASSO) [55], support vector regression (SVR) [56], and random forests (RF) [57] were employed as the regression methods. As descriptors for these regression methods, a generalized set of chemical element-based descriptors (132 descriptors) [2] obtained by the matminer [58] Python library were used. For the SVR, a radial basis function (RBF) kernel was used. In addition, SOAP kernel [9, 10] was employed as a representative structural descriptor-based predictive model. DScribe package [59] was used to calculate the SOAP of each atom and construct SOAP kernels formed by a pair of local environments. Ridge regression [60] with SOAP kernel descriptors was used to predict the properties. The sizes of training+validation and test datasets were 625 and 125, respectively. For training and validation data, 5-fold cross validations were used to optimize the hyperparameters relevant to the minimization of overfitting for the regression methods.

## 3. Results and Discussion

### 3.1. CGCNN

Herein, before describing the prediction results, we note that the names of the pretrained and target models are written based on the size of the training dataset and target property. For instance, 500-NM-$E_g$ indicates that the target property of the predictive model is $E_g$, the size of the training dataset is 500 (sizes of both validation and test datasets are 125), and only NM data are included in the datasets.

Table 1 summarizes material-property datasets that are used in this study. Most of the materials are bulk-type inorganic materials, and their crystal structures and properties are loaded from the MPD. A mean absolute difference of the data indicates a baseline for a predictive model.

**Table 1.** Summary of the material-property datasets used in this study.

| Material Property | Size of dataset | Minimum value | Maximum value | Mean value | Mean absolute difference | Data source Ref. |
|---|---|---|---|---|---|---|
| $E_g$ | 59923 | 0.100 eV | 9.721 eV | 2.179 eV | 1.256 eV | [45] |
| $\Delta E_f$ | 118286 | −4.612 eV/atom | 5.124 eV/atom | −1.481 eV/atom | 0.968 eV/atom | [45] |
| $\Delta E_f$ (NM) | 59923 | −4.612 eV/atom | 3.797 eV/atom | −2.021 eV/atom | 0.768 eV/atom | [45] |
| $K_{VRH}$ [a,b] | 12990 | 0.301 | 2.639 | 1.874 | 0.304 | [45] |
| $\varepsilon_r$ [a] | 4182 | 0.063 | 2.992 | 1.096 | 0.265 | [45] |
| GW-$E_g$ | 270 | 0.357 eV | 14.548 eV | 4.996 eV | 2.486 eV | [39] |

[a] The values are in $\log_{10}$.

[b] The unit in $\log_{10}$ is GPa.

All the data were separated into NM and metallic data using $E_g$ with a threshold value of 0.1 eV. However, when we consider constituent chemical elements in the materials, NM (semiconductors and insulators) materials, such as oxides, are mainly inorganic compounds formed by combining cationic and anionic elements, whereas metallic materials, such as transition metal alloys, are mainly alloys consisting of combinations of metallic elements. For instance, among 59923 NM materials, 52089 are oxides, nitrides, and halogenides. The NM materials mainly have ionic or covalent bonds, whereas the metallic materials mainly have metallic bonds. Therefore, we consider that the atom and bond features saved in the neural network architecture may be different from each other.

Figure 3 shows scatter plots between the $\Delta E_f$ and $E_g$ properties of 118286 materials. The scatter plots indicate that there is no strong correlation between $\Delta E_f$ and $E_g$. The linear correlation coefficients ($r_p$) between the two properties for the entire data and NM data are −0.49 and −0.33, respectively. For the 59923 NM materials, the MAE of the $\Delta E_f$ and $E_g$ obtained via an ordinary linear regression using $E_g$ and $\Delta E_f$ were 0.72 eV/atom and 1.17 eV, respectively. This implies that more complex regression methods and descriptor sets are necessary for their predictions.

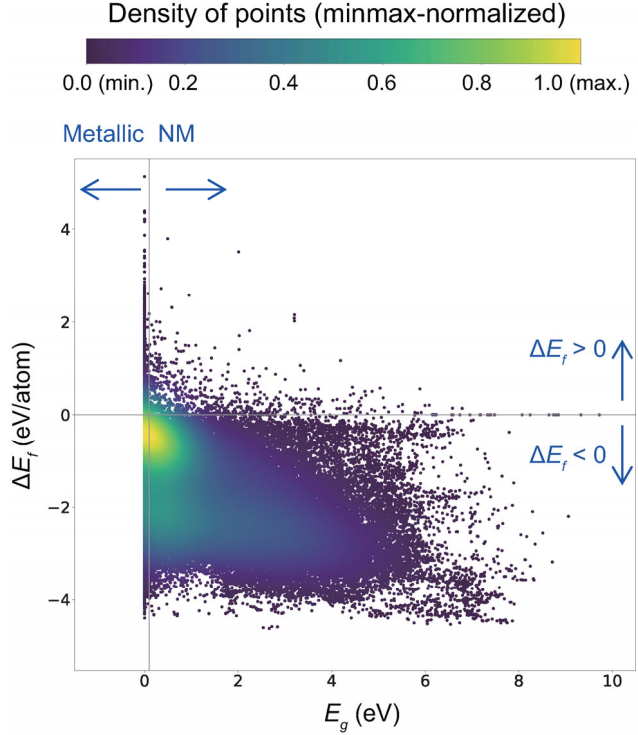

**Fig. 3.** Scatter plots between the $\Delta E_f$ and $E_g$ quantities of 118286 materials. The color scale represents the relative density of points generated via Gaussian kernel density estimation [61]. The probability density function is minmax-normalized, and the Gaussian bandwidth is three times the Scott's factor [61].

Figure 4 shows the dependence of the prediction error of the test data by CGCNN for $E_g$ and $\Delta E_f$ on the size of training dataset. The prediction errors decrease with an increase in the size of training data. When the sizes of training datasets are 54k and 113k, the prediction errors for $E_g$ and $\Delta E_f$ are 0.358 eV and 0.046 eV/atom, respectively. The prediction error of $\Delta E_f$ is comparable to the accuracy of the density functional theory (DFT) method with respect to the experiments [31]. We expect that the prediction error can further decrease when we collect more training data. This is because the decreasing tendency is not saturated yet. In addition, both prediction error and size of training dataset in $\log_{10}$ scale shows a strong linear relationship, as reported in previous ML models for other systems [62, 63]. The $r_p$ of $E_g$, $\Delta E_f$, and NM-$\Delta E_f$ vs. size of training dataset in $\log_{10}$ scale are −1.000, −0.999, and −1.000, respectively.

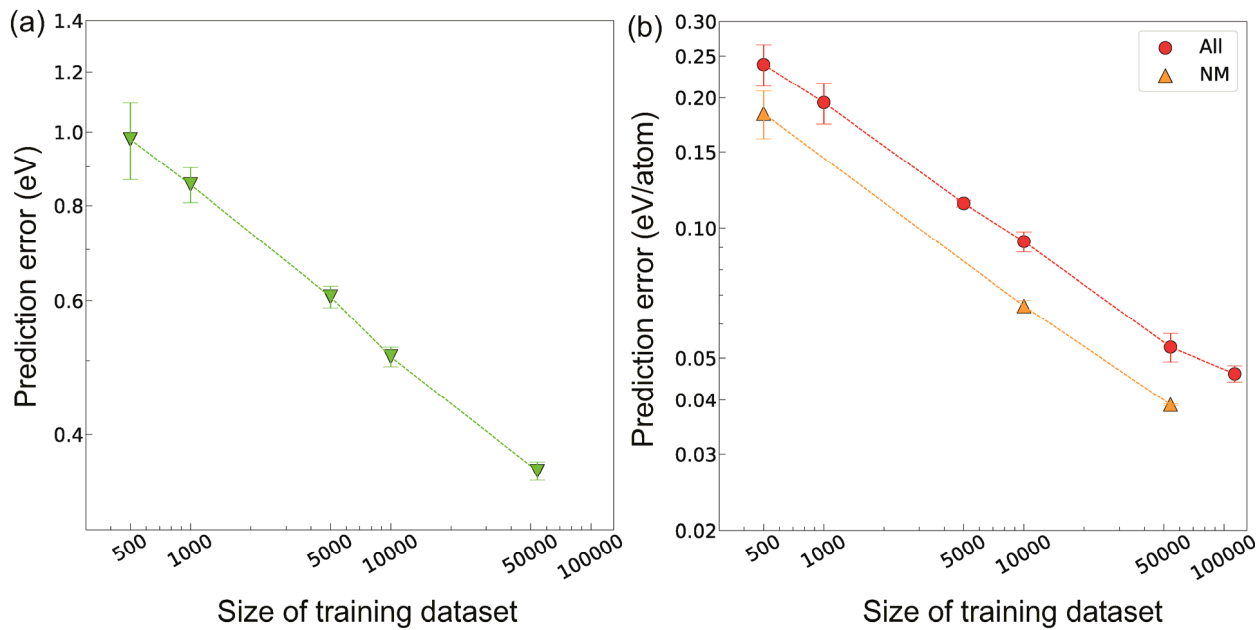

**Fig. 4.** Dependence of prediction error (MAE of test data) for (a) $E_g$ and (b) $\Delta E_f$ on size of the training dataset by CGCNN. Horizontal and vertical axes are in $\log_{10}$ scale. Error bars indicate one standard deviation for prediction errors from several predictive models with different data samplings.

As mentioned above, because the bond natures and constituent elements of the NM and metallic materials are different, predictive models for $\Delta E_f$ with only NM material data were separately constructed. The prediction error with only NM material data is smaller than that with metallic and NM material data combined when the same size of training dataset is used.

*3.2. CGCNN-based TL-CGCNN*

Predictive models usually suffer from unsatisfactory accuracy owing to a small amount of data. Therefore, we intentionally prepared target models with a small amount of data for applying a TL-CGCNN to $E_g$ and $\Delta E_f$. The results are shown in Fig. 5, and numerical values are summarized in Supplementary Table 2. The pretrained models with a large amount of data were used for predicting targets of counterpart properties.

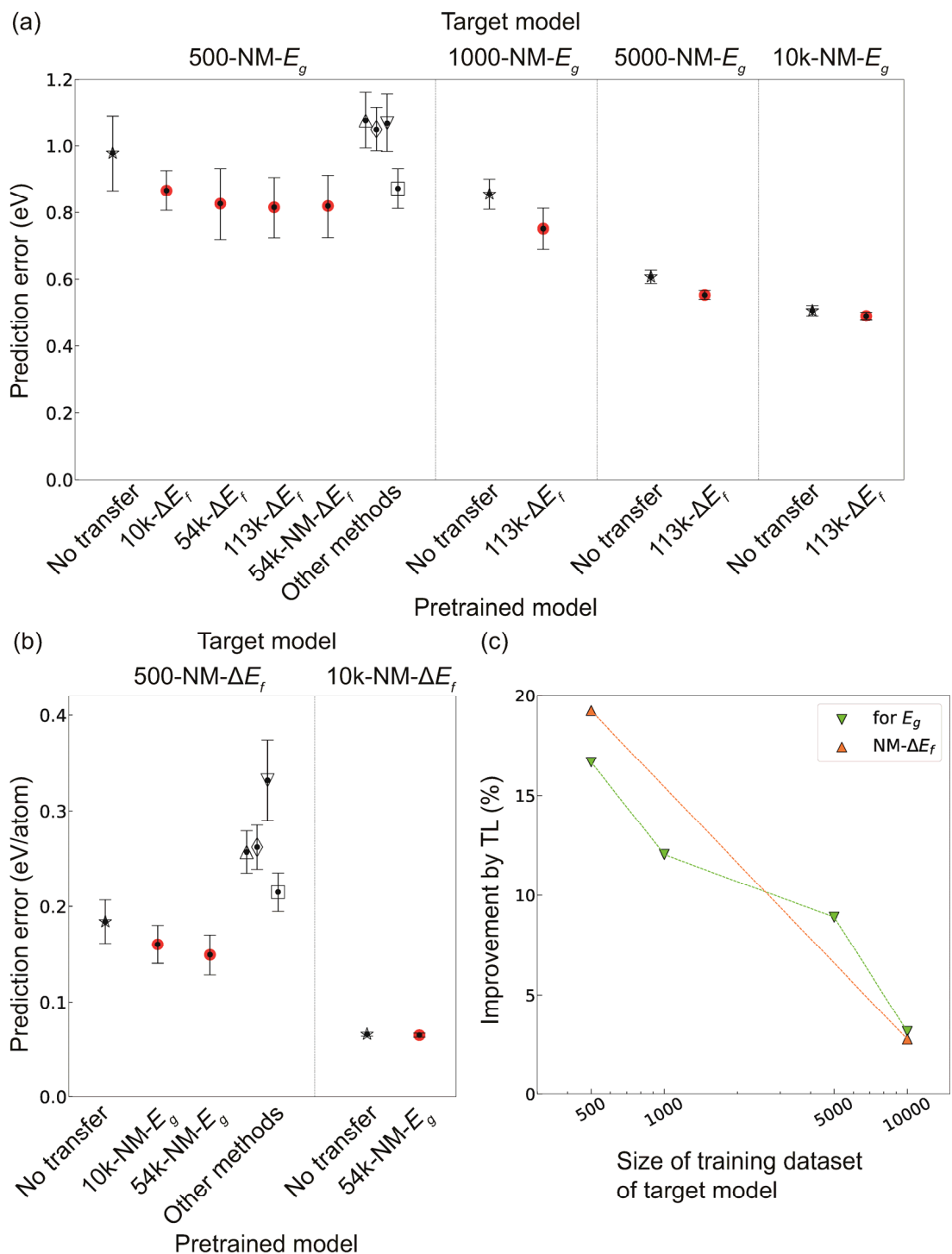

**Fig. 5.** Comparison of prediction error for (a) $E_g$ and (b) NM-$\Delta E_f$. The predictions via the CGCNN (no transfer) and TL-CGCNN are displayed as open star and red circle symbols, respectively. The predictions via other regression methods, namely, PLS, LASSO, SVR, and RF, with elemental descriptors are displayed as open up-pointing triangle, diamond, down-pointing triangle, and square symbols, respectively. Error bars indicate one standard deviation for prediction errors from ten predictive models with different data samplings. The baselines, obtained as the mean absolute differences of the data distribution, for $E_g$ and NM-$\Delta E_f$ are 1.26 and 0.77 (in the same unit as in each subfigure), respectively. (c) Dependence of percentage improvement of prediction error by best TL-CGCNN with respect to plain CGCNN on size of training dataset of target model.

First, Fig. 5(a) shows a comparison of the predictive performances for $E_g$. When the size of the training dataset is 500, the prediction error of $E_g$ by the CGCNN without TL is 0.978 eV. Using TL-CGCNN with the 10k-$\Delta E_f$, 54k-$\Delta E_f$, and 113k-$\Delta E_f$ pretrained models, the prediction error decreases to 0.866, 0.826, and 0.815 eV, respectively. The more training data we use for the pretrained models, the more the prediction powers are improved.

In addition, we performed a statistical significance analysis to judge whether the decrease in prediction error via TL is meaningful or accidental. For this analysis, the paired *t*-test was employed. To confirm that the mean averages of the two distributions are different, the *p*-value should be lower than 0.01 (threshold for 99% significance level). The *p*-value of the paired *t*-test using ten MAE values from the CGCNN and TL-CGCNN with the 113k-$\Delta E_f$ pretrained model is $6.9 \times 10^{-5}$. Therefore, we conclude that the TL-CGCNN exhibits better performance than the CGCNN with a 99% significance level for this prediction.

Second, we investigated predictive performances for $\Delta E_f$. The prediction error of $\Delta E_f$ by the CGCNN without TL is 0.239 eV/atom when the size of the training dataset is 500. When we use TL-CGCNNs with the 10k-NM-$E_g$ and 54k-NM-$E_g$ pretrained models, the prediction error decreases to 0.236 and 0.222 eV/atom, respectively. However, despite the small decrease in prediction error, we conclude that an improvement via the TL is not achieved. This is because the *p*-value of the statistical significance analysis between the CGCNN and TL-CGCNN with the 54k-NM-$E_g$ pretrained model is too large (0.15).

Additionally, Fig. 5(b) shows a comparison of the predictive performances for $\Delta E_f$, but with only NM material data (NM-$\Delta E_f$). When the target model is 500-NM-$\Delta E_f$, the prediction error by the TL-CGCNN with the 54k-NM-$E_g$ pretrained model is 0.149 eV/atom, which is much smaller than that by the CGCNN without TL (0.184 eV/atom). Contrary to the former predictive models ($\Delta E_f$ of all data), the *p*-value of the statistical significance analysis between the CGCNN and TL-CGCNN with the 54k-NM-$E_g$ pretrained model is small ($4 \times 10^{-4}$). Therefore, we conclude that the TL-CGCNN exhibits better performance than the CGCNN with a 99% significance level for this prediction. In this case, the data of both pretrained and target models were NM data. The bond features are mainly formed by ionic or covalent bonds, and the crystal graphs are formed by combinations of cations and anions. We presume that this pretrained information could be smoothly transformed into the predictions for the NM material data, but not for all data, such as the data for metallic materials, which are less overlapped with the pretrained data in the descriptor space.

Subsequently, we tested the dependence of the TL-CGCNN on the size of dataset of the target models. Using the 113k-$\Delta E_f$ pretrained model, the prediction errors are 0.866, 0.752, 0.553, and 0.490 eV when the target models are 500-$E_g$, 1000-$E_g$, 5000-$E_g$, and 10k-$E_g$, respectively. As shown in Fig. 5(c), the percentage improvements in prediction error for the case without TL tend to decrease by 16.7, 12.0, 9.0 and 3.2%, respectively. This was also tested for the prediction of $\Delta E_f$ for the NM data. Using the 54k-NM-$E_g$ pretrained model, the prediction errors are 0.149 and 0.065 eV/atom (19.2 and 2.8% improvement in predictions without the TL-CGCNN) when the target models are 500-NM-$\Delta E_f$ and 10k-NM-$\Delta E_f$, respectively. This implies that using the TL-CGCNN is more efficient when we have less data.

We confirmed that the TL-CGCNN for $E_g$ and $\Delta E_f$ with pretrained models based on the counterpart properties improved the predictions despite a weak correlation, as shown in Fig. 3. This implies that the atom and bond features from crystal graphs captured in the convolutional layers in the pretrained model with larger data are flexibly adapted to predict other material properties.

*3.3. TL-CGCNN for computationally demanding properties*

Additionally, we tested the TL-CGCNN for predictive models for other properties, namely, $K_{VRH}$, $\varepsilon_r$, and GW-$E_g$. Because these properties are more computationally expensive, there are much less data than $E_g$ or $\Delta E_f$ data. Considering their broad distributions, the $K_{VRH}$ and $\varepsilon_r$ values in a $\log_{10}$ scale were used. The unit of the $K_{VRH}$ value inside the $\log_{10}$ is GPa.

Figure 6 shows the prediction errors of three properties via the CGCNN without TL and TL-CGCNN. The prediction error of the 500-$K_{VRH}$ by the CGCNN without TL is 0.123. When the 113k-$\Delta E_f$ pretrained model is used, the prediction error decreases to 0.112 (8.7%). The prediction error of the 500-NM-$\varepsilon_r$ by the CGCNN without TL is 0.181. It decreases to 0.169 (7.1%) and 0.163 (10.0%) when the 113k-$\Delta E_f$ and 54k-NM-$E_g$ pretrained models are employed, respectively. Despite the additional training data of the 113k-$\Delta E_f$ pretrained model, a smaller prediction error of the 54k-NM-$E_g$ pretrained model-based prediction is obtained. For 4182 NM materials that have $\varepsilon_r$ data, this may be ascribed to the correlation (−0.48) between $E_g$ and $\varepsilon_r$ (in $\log_{10}$ scale) being much stronger than the correlation (0.07) between $\Delta E_f$ and $\varepsilon_r$ (in $\log_{10}$ scale). The prediction error of the 180-NM-GW-$E_g$ by the CGCNN without TL is 0.783 eV. The smallest prediction error by the TL-CGCNN among various pretrained models is achieved by the 113k-$\Delta E_f$ pretrained model, whose value decreases to 0.591 eV (24.5%). For the 500-$K_{\text{VRH}}$, 500-NM-$\varepsilon_r$, and 180-NM-GW-$E_g$ predictions, the $p$-values obtained via the statistical significance analysis between the CGCNN without TL and corresponding best TL-CGCNN models are 0.002, 0.004, and $2\times10^{-4}$, respectively. Therefore, we conclude that the TL-CGCNN shows better performance than the CGCNN without TL for these three predictions with a 99% significance level.

When more data are used for the prediction of $K_{\text{VRH}}$ and $\varepsilon_r$, (8194 and 2788 training data for $K_{\text{VRH}}$ and $\varepsilon_r$, respectively), we obtain a decreased prediction error by the TL-CGCNN. However, the improvement via TL tends to decrease compared to when the target model has less data as shown in Fig. 6(d). The percentage improvements in $K_{\text{VRH}}$ and $\varepsilon_r$ using TL are 1.7 and 3.7%, respectively. This implies that the TL-CGCNN may be more useful for predictions of target properties that suffer from small amount of data.

Table 2 shows equivalent size of training dataset for CGCNN to achieve the same prediction error by corresponding best TL-CGCNN with 500 training data of target model. This is obtained by an interpolation using the linear relationship between $\log_{10}$(prediction error) and $\log_{10}$(size of dataset) as shown in Fig. 4(b). The equivalent sizes of training dataset for CGCNN to achieve the same prediction error by corresponding best TL-CGCNN become approximately twice. Accordingly, these results demonstrate that using the TL-CGCNN is useful for such practical properties that usually suffer from limited size of dataset.

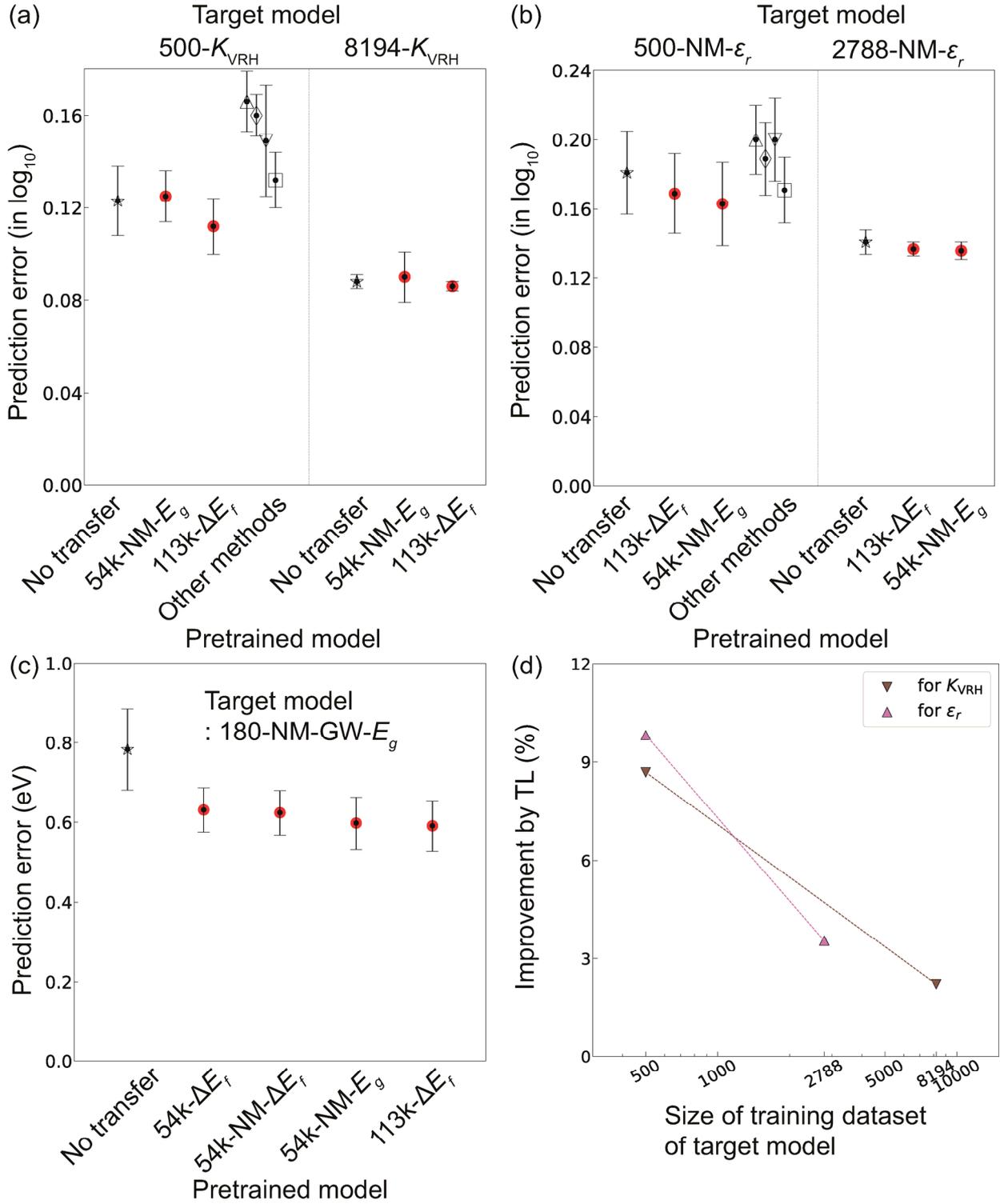

**Fig. 6.** Comparison of prediction errors of (a) $K_{VRH}$, (b) $\varepsilon_r$, and (c) GW-$E_g$. The predictions via the CGCNN (no transfer) and TL-CGCNN are displayed as open star and red circle symbols, respectively. The predictions via other regression methods, namely, PLS, LASSO, SVR, and RF, with elemental descriptors are displayed as open up-pointing triangle, diamond, down-pointing triangle, and square symbols, respectively. The unit of $K_{VRH}$ in $\log_{10}$ is GPa. Error bars indicate one standard deviation for prediction errors from ten predictive models with different data samplings. The baselines, obtained as the mean absolute differences of the data distribution, for $K_{VRH}$, $\varepsilon_r$, and GW-$E_g$ are 0.30, 0.27, and 2.49 (in the same unit as in each subfigure), respectively. (d) Dependence of percentage improvement of prediction error by best TL-CGCNN with respect to plain CGCNN on size of training dataset of target model.

**Table 2.** Equivalent size of training dataset for CGCNN to achieve the same prediction error obtained by corresponding best TL-CGCNN when the size of the training dataset of target model is 500.

| Target model | 500-$E_g$ | 500-NM-$\Delta E_f$ | 500-$K_{VRH}$ | 500-NM-$\varepsilon_r$ |
| --- | --- | --- | --- | --- |
| Equivalent size of training dataset | 1200 | 925 | 1068 | 1028 |

*3.4. Additional issues*

We also tested predictions of 500-NM-$E_g$, 500-NM-$\Delta E_f$, 500-$K_{VRH}$, and 500-NM-$\varepsilon_r$ via PLS [54] LASSO [55], SVR [56], and RF [57]. These methods were supervised using a generalized set of chemical element-based descriptors [2]. The results are shown in Fig. 5 and 6. In addition, there are (similarity) kernel-based models [9, 10, 63-65] that use structural descriptors based on information of atomic coordinates. In this study, SOAP kernel [9, 10] with ridge regression [60] was used to compare the predictions. The numerical values of the predictive performances are summarized in Supplementary Table S2.

Among the four regression methods with the chemical element-based descriptors, RF exhibits the best performance. The CGCNN exhibits better performance in the predictions of 500-NM-$\Delta E_f$ and 500-$K_{VRH}$, but poorer performance in the predictions of 500-NM-$E_g$ and 500-NM-$\varepsilon_r$ than the RF. However, the TL-CGCNN achieves the smallest prediction errors for the four predictions. The predictions with other widely used regression methods such as RF can be improved with descriptors that are more specialized and have strong correlations with the target properties; however, it is not easy to find additional useful descriptors for each target property.

Additionally, as a descriptor of local structure, we employed SOAP that has been extensively used for regression models or generating ML potentials [66, 67]. For organic molecules [10, 65, 68] or nanoclusters [69-71], the target properties can be efficiently supervised by structural descriptors such as bond lengths and angles. This is because most of the data consist of limited types of chemical elements such as carbon, hydrogen, oxygen, and nitrogen. On the contrary, the MPD data used in this study comprise various inorganic materials; therefore, most of the chemical elements in the periodic table participate in the data. Notably, as summarized in Supplementary Table S2, the SOAP kernel did not show better performance compared to the TL-CGCNN. The prediction errors by the SOAP kernel for predictions of 500-NM-$E_g$, 500-NM-$\Delta E_f$, 500-$K_{VRH}$, and 500-NM-$\varepsilon_r$ are larger compared to those by the RF with the chemical element-based descriptors. This indicates that there is little similarity in local environments among chemical elements in such a small number (five hundred) of data samples selected from a wide range of inorganic materials. Karamad *et al*. [18] reported that SOAP descriptors with kernel ridge regression showed similar performance to that of the CGCNN for predictions of $\Delta E_f$ and $E_g$ of inorganic materials using more than 26k data samples from MPD.

Therefore, we emphasize that the TL-CGCNN is a powerful and flexible prediction tool with merits for a small amount of data. This tool facilitates the construction of pretrained models with big data, and it can systematically capture features of both chemical elements and crystal structures.

Deep neural networks such as CGCNN have many hyperparameters. For an effective TL, we need to carefully select hyperparameters based on the following considerations: transferred information is to be conserved and flexibly used for the target predictive model, and overfitting is to be reduced. Based on these considerations, we optimized the learning rate and the length of the vector of the FC layer for a case study. The results are described in Supplementary Chapter S1.

Fine-tuning has been used as the TL-CGCNN method in this study. Additionally, we tested the layer freezing method. We only optimized the parameters of the last FC layer, freezing those of the earlier layers. As a result, for the investigated prediction targets, fine-tuning exhibits better predictions. The detailed results are described in Supplementary Chapter S2. In some previous studies using deep convolutional neural networks with many multiple layers [23, 33], the concept of mixing fine-tuning and layer freezing has been suggested, that is, only early convolutional layers are frozen and the other layers including convolutional layers are optimized.

## 4. Conclusion

In this study, TL-CGCNNs were applied to the predictive models of various material properties, which suffer from inaccuracy owing to small datasets of target properties. We quantitatively confirmed that the predictions can be improved using the TL-CGCNN with the pretrained model constructed with big data. In addition, we quantitatively observed that the prediction of properties in target models via TL-CGCNN becomes more accurate when the pretrained model is trained with bigger data and the pretrained and target models are more strongly correlated. Because the problem of an insufficient amount of data arises in many MI issues, we expect that a combination of general descriptors such as CGCNN and TL can be more widely employed.

**Acknowledgement**

J. L. and R. A. thank Xie and Grossman in Massachusetts Institute of Technology (MIT) for an open source code of CGCNN [13] and Editage (www.editage.com) for English language editing.

**Supplementary Information**

*S1. Optimization of hyperparameters for TL-CGCNN*

Deep neural networks such as crystal graph convolutional neural network (CGCNN) have many hyperparameters. For strict optimization, all the hyperparameters must be optimized. However, it requires heavy computations to optimize such a large set of hyperparameters. In this study, we tested the influence of the learning rate and length of feature vector of the fully-connected (FC) layer on the prediction error using the CGCNN with transfer learning (TL-CGCNN) while fixing the hyperparameters summarized in Supplementary Table S1. We used a case study of a predictive model of the 500-NM-$E_g$ by the TL-CGCNN with the 113k-$\Delta E_f$ pretrained model.

The learning rate for optimizing weight and bias is determined using an initial learning rate and learning rate decay. At early epochs, the initial learning rate becomes the learning rate. Thereafter, the learning rate can be decayed by a multiplicative factor at later epochs. In addition, there can be various combinations relevant to modifying the learning rate schedule, such as the frequency of decay and the epochs at which decay occurs. In this study, we only check the initial learning rate and multiplicative factor of learning rate decay.

Figure S1(a) shows the prediction error of the 500-NM-$E_g$ depending on the initial learning rate. The initial learning rate of the pretrained model is 0.01. When the initial learning rate is too large (1), the prediction errors are larger. This may be because the optimized parameters loaded from the pretrained models may disappear. The smallest prediction error is obtained when the initial learning rate is 0.005. Therefore, for the "fine" tuning, it is recommended to use an adequately small learning rate to avoid losing the transferred knowledge. However, when the learning rate is too small (0.0001), the prediction error increases again. This indicates that effective learning was not performed.

Figure S1(b) shows the prediction error of the 500-NM-$E_g$ depending on the multiplicative factor of learning rate decay. The multiplicative factor of learning rate decay of the pretrained model is 0.1. The smallest prediction error is obtained when the multiplicative factor of learning rate decay is 0.0001. In the investigated range of 0.0001–10 for the multiplicative factor of learning rate decay, the differences in the prediction errors are not as large compared to those depending on the change in the initial learning rate. This may be because conservation of the transferred knowledge strongly depends on the initial learning rate at early epochs for this predictive model.

Figure S1(c) shows the prediction error of the 500-NM-$E_g$ depending on the length of feature vector of the FC layer of the target model. In the pretrained model, the length of the feature vector of the FC layer is 128. The smallest prediction error is obtained when the length of feature vector of the FC layer is 64. The prediction errors are similar to the values ranging between 0.815–0.836 eV when the lengths of feature vector of the FC layer ranging between 4–256. However, we recommend not using too small or too large length of the feature vector of the FC layer because it can result in much larger prediction errors.

**Table S1.** Main hyperparameters used for the CGCNN and TL-CGCNN.

| Hyperparameters | Value |
|---|---|
| Minibatch size | 256 |
| Length of atom feature vector $v_i$ after an embedding layer | 64 |
| Number of convolutional layers | 3 |
| Length of feature vector of FC layer | 128→64 (in case of transfer learning) |
| Number of FC layer | 1 |
| Optimizer for loss function | Stochastic gradient descent |
| Initial learning rate | 0.01→0.005 (in case of transfer learning) |
| Multiplicative factor of learning rate decay [a] | 0.1 |
| Dropout | 0 |

[a] In the CGCNN code, the learning rate decay occurs at the 100$^{th}$ epoch as a default.

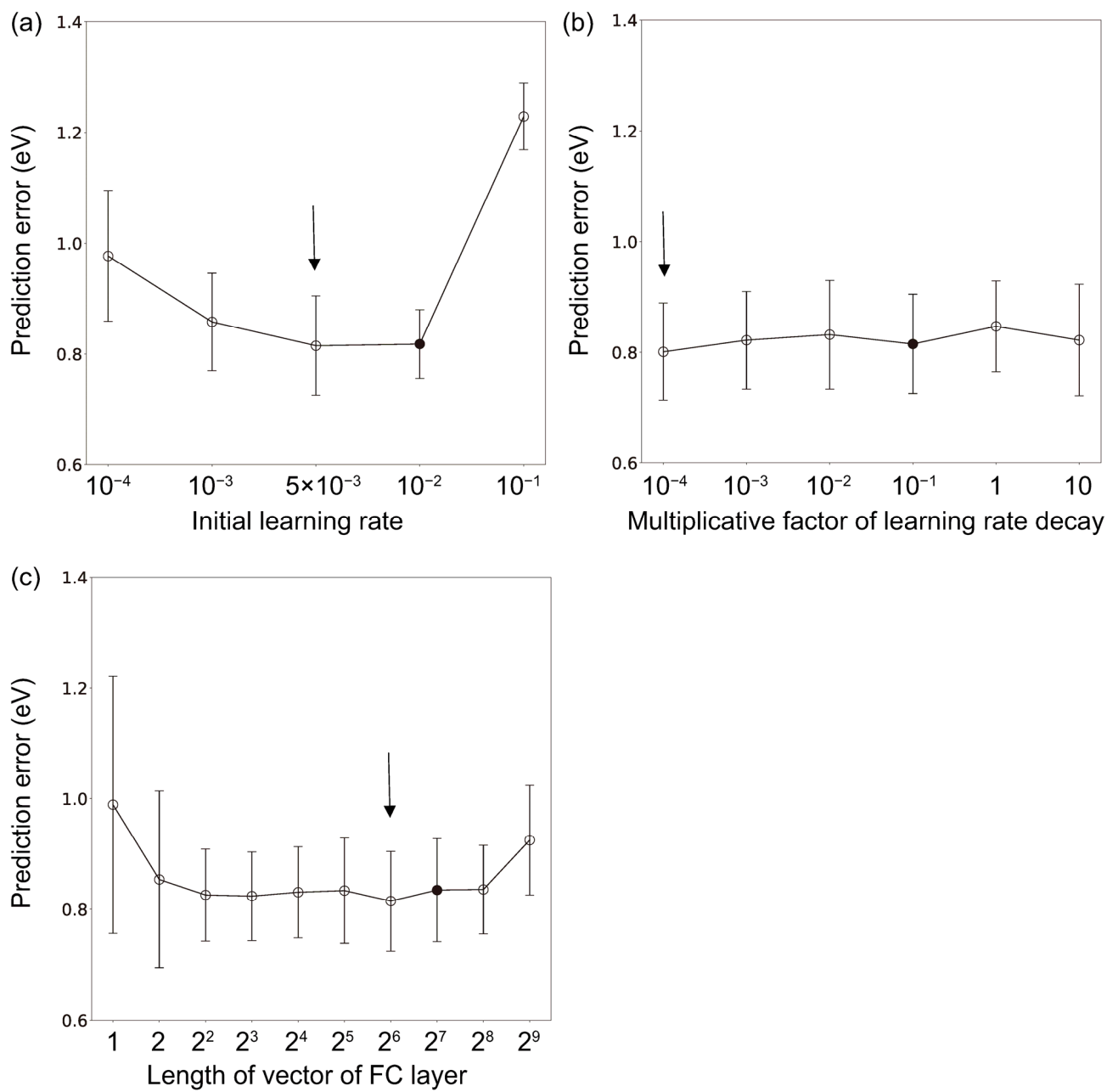

**Figure S1.** Dependences of prediction error of TL on hyperparameters. Prediction error of the 500-NM-$E_g$ prediction model depending on (a) initial learning rates, (b) learning rate decays, and (c) lengths of feature vector of the FC layer by TL with the 113k-$\Delta E_f$ pretrained model. Closed marks indicate the hyperparameter used in the pretrained model. A vertical indicator shows the optimal hyperparameter with the smallest prediction error. Error bars indicate one standard deviation for prediction errors from ten predictive models with different data samplings.

*S2. Test of layer freezing*

In this study, fine-tuning was used as the TL-CGCNN. For comparison, layer freezing is also investigated as the TL-CGCNN. We only optimized the weights and biases of the last FC layer by freezing those of the other earlier layers. For the prediction targets, the 500-NM-$E_g$, 500-NM-$\Delta E_f$, 500-$K_{VRH}$, 500-NM-$\varepsilon_r$, and 180-NM-GW-$E_g$ were used with the 113k-$\Delta E_f$, 54k-NM-$E_g$, 113k-$\Delta E_f$, 54k-NM-$E_g$, and 113k-$\Delta E_f$ pretrained models, respectively, which showed the best performance when fine-tuning was used. The identical hyperparameters were used for the two methods. Supplementary Table S2 summarizes the comparison of the prediction errors via the CGCNN without TL, fine-tuning, layer freezing, and other regression methods. Additionally, the prediction errors of the training data are listed.

For all five predictive models using the TL-CGCNN, fine-tuning shows smaller prediction errors than layer freezing. For the 500-NM-$E_g$ target model, layer freezing exhibits a prediction error of 0.956 eV, which is slightly smaller than that of the prediction without the TL-CGCNN with a prediction error of 0.978 eV. However, for the 500-NM-$\Delta E_f$, 500-$K_{VRH}$, 500-NM-$\varepsilon_r$, and 180-NM-GW-$E_g$ target models, layer freezing exhibits worse predictions compared to those without the TL.

Although the prediction improvements are not achieved via layer freezing, the prediction error differences between the training data and test data are much smaller compared to those obtained via fine-tuning. This implies that overfitting is reduced. However, relatively larger errors in the training data obtained via layer freezing indicate that the freezing of parameters of the convolutional layers in the CGCNN obstructs the optimization of the predictive model to a satisfactory level. In previous studies using deep convolutional neural networks with many multiple layers [1, 2], it was suggested to mix the concept of the two methods, that is, only early convolutional layers are frozen and the other layers including convolutional layers are optimized.

**Table S2.** Comparison of prediction errors (as MAE) of training and test data via the CGCNN without TL, fine-tuning, layer freezing, and other regression methods. The bold model emphasizes the smallest error of the test data. The value in the parentheses is the standard deviation of MAE of ten trials. Different training/validation/test data are used for each trial.

| Target model | Pretrained model | Method | Prediction error of training data | Prediction error of test data [c] | Unit |
|---|---|---|---|---|---|
| 500-NM-$E_g$ | | CGCNN | 0.576 (0.129) | 0.978 (0.113) | eV |
| **500-NM-$E_g$** | **113k-$\Delta E_f$** | **Fine tuning by TL-CGCNN** | **0.511 (0.144)** | **0.815 (0.090)** | **eV** |
| 500-NM-$E_g$ | 113k-$\Delta E_f$ | Layer freezing by TL-CGCNN | 0.919 (0.046) | 0.956 (0.099) | eV |
| 500-NM-$E_g$ | | PLS with elemental descriptors | 0.940 (0.021) | 1.051 (0.066) | eV |
| 500-NM-$E_g$ | | LASSO with elemental descriptors | 0.850 (0.021) | 1.078 (0.084) | eV |
| 500-NM-$E_g$ | | SVR with elemental descriptors | 0.720 (0.033) | 1.070 (0.087) | eV |
| 500-NM-$E_g$ | | RF with elemental descriptors | 0.367 (0.038) | 0.872 (0.060) | eV |
| 500-NM-$E_g$ | | Ridge with SOAP kernel | 1.124 (0.017) | 1.130 (0.059) | eV |
| 500-NM-$\Delta E_f$ | | CGCNN | 0.081 (0.022) | 0.184 (0.023) | eV/atom |
| **500-NM-$\Delta E_f$** | **54k-NM-$E_g$** | **Fine tuning by TL-CGCNN** | **0.059 (0.006)** | **0.149 (0.021)** | **eV/atom** |
| 500-NM-$\Delta E_f$ | 54k-NM-$E_g$ | Layer freezing by TL-CGCNN | 0.371 (0.018) | 0.424 (0.033) | eV/atom |
| 500-NM-$\Delta E_f$ | | PLS with elemental descriptors | 0.211 (0.010) | 0.257 (0.022) | eV/atom |
| 500-NM-$\Delta E_f$ | | LASSO with elemental descriptors | 0.237 (0.025) | 0.262 (0.023) | eV/atom |
| 500-NM-$\Delta E_f$ | | SVR with elemental descriptors | 0.222 (0.045) | 0.332 (0.042) | eV/atom |
| 500-NM-$\Delta E_f$ | | RF with elemental descriptors | 0.093 (0.004) | 0.215 (0.020) | eV/atom |
| 500-NM-$\Delta E_f$ | | Ridge with SOAP kernel | 0.469 (0.024) | 0.476 (0.025) | eV/atom |
| 500-$K_{VRH}$ | | CGCNN | 0.067 (0.020) | 0.123 (0.015) | [a,b] |
| **500-$K_{VRH}$** | **113k-$\Delta E_f$** | **Fine tuning by TL-CGCNN** | **0.075 (0.022)** | **0.112 (0.012)** | **[a,b]** |
| 500-$K_{VRH}$ | 113k-$\Delta E_f$ | Layer freezing by TL-CGCNN | 0.149 (0.008) | 0.161 (0.011) | [a,b] |
| 500-$K_{VRH}$ | | PLS with elemental descriptors | 0.132 (0.008) | 0.166 (0.013) | [a,b] |
| 500-$K_{VRH}$ | | LASSO with elemental descriptors | 0.145 (0.007) | 0.160 (0.009) | [a,b] |
| 500-$K_{VRH}$ | | SVR with elemental descriptors | 0.075 (0.039) | 0.149 (0.024) | [a,b] |
| 500-$K_{VRH}$ | | RF with elemental descriptors | 0.059 (0.008) | 0.132 (0.012) | [a,b] |
| 500-$K_{VRH}$ | | Ridge with SOAP kernel | 0.219 (0.010) | 0.234 (0.020) | [a,b] |
| 500-NM-$\varepsilon_r$ | | CGCNN | 0.120 (0.032) | 0.181 (0.024) | [a] |
| **500-NM-$\varepsilon_r$** | **54k-NM-$E_g$** | **Fine tuning by TL-CGCNN** | **0.095 (0.016)** | **0.163 (0.024)** | **[a]** |
| 500-NM-$\varepsilon_r$ | 54k-NM-$E_g$ | Layer freezing by TL-CGCNN | 0.181 (0.008) | 0.193 (0.032) | [a] |
| 500-NM-$\varepsilon_r$ | | PLS with elemental descriptors | 0.165 (0.006) | 0.200 (0.020) | [a] |
| 500-NM-$\varepsilon_r$ | | LASSO with elemental descriptors | 0.175 (0.005) | 0.189 (0.021) | [a] |
| 500-NM-$\varepsilon_r$ | | SVR with elemental descriptors | 0.074 (0.031) | 0.200 (0.024) | [a] |
| 500-NM-$\varepsilon_r$ | | RF with elemental descriptors | 0.079 (0.010) | 0.171 (0.019) | [a] |
| 500-NM-$\varepsilon_r$ | | Ridge with SOAP kernel | 0.224 (0.007) | 0.225 (0.025) | [a] |
| 180-NM-GW-$E_g$ | | CGCNN | 0.302 (0.140) | 0.783 (0.102) | eV |
| **180-NM-GW-$E_g$** | **113k-$\Delta E_f$** | **Fine tuning by TL-CGCNN** | **0.196 (0.082)** | **0.591 (0.063)** | **eV** |
| 180-NM-GW-$E_g$ | 113k-$\Delta E_f$ | Layer freezing by TL-CGCNN | 0.525 (0.178) | 0.867 (0.152) | eV |

[a] The values are in $\log_{10}$.
[b] The unit in $\log_{10}$ is GPa.
[c] Note that the amount of test data is 25% of the training data.

*S3. Effective U values for GGA+U calculations*

In Materials Project Database (MPD) [3], effective U values were consistently used for *d*-orbitals of inorganic compounds including transition metals. Table S3 summarizes an effective U value for each transition metal.

**Table S3.** U values for *d*-orbitals in the GGA+U calculations used in MPD.

| Transition metal | Effective U | Transition metal | Effective U | Transition metal | Effective U |
| --- | --- | --- | --- | --- | --- |
| Sc | 0.0  | Y  | 0.0  | La | 0.0 |
| Ti | 0.0  | Zr | 0.0  | Hf | 0.0 |
| V  | 3.25 | Nb | 0.0  | Ta | 0.0 |
| Cr | 3.7  | Mo | 4.38 | W  | 6.2 |
| Mn | 3.9  | Tc | 0.0  | Re | 0.0 |
| Fe | 5.3  | Ru | 0.0  | Os | 0.0 |
| Co | 3.32 | Rh | 0.0  | Ir | 0.0 |
| Ni | 6.2  | Pd | 0.0  | Pt | 0.0 |
| Cu | 0.0  | Ag | 0.0  | Au | 0.0 |
| Zn | 0.0  | Cd | 0.0  | Hg | 0.0 |

**Supplementary References**